\documentclass[preprint,12pt,authoryear]{elsarticle}

\usepackage{amssymb}

\begin{document}

\begin{frontmatter}

\title{Are we visible to advanced alien civilizations?}

\author{Z.N. Osmanov}

\affiliation{organization={School of Physics, Free University of Tbilisi},
            city={Tbilisi},
            postcode={0183}, 
            country={Georgia}}

\affiliation{organization={E. Kharadze Georgian National Astrophysical Observatory},
            city={Abastumani},
            postcode={0301}, 
            country={Georgia}}

\begin{abstract}
We considered the question of how our artificial constructions are visible to  advanced extraterrestrial civilizations. Taking the universality of the laws of physics, we found that the maximum distance where the detection is possible is of the order of $3000$ ly and under certain conditions Type-II advanced alien societies might be able to resolve this problem.
\end{abstract}




\end{frontmatter}


\section{Introduction}

In the context of the search for extraterrestrial (ET) intelligence (SETI), a closely related question can be formulated: If they exist, are we visible to them? Answering this question inevitably implies consideration of the technological level to which an alien civilization belongs.

In this context, it is worth noting that \cite{kardashev} has introduced a technological classification of advanced alien societies. According to his approach, there might be three major technological civilizations: Type-I is an alien society that consumes the total energy incident on a planet from their host star; Type-II is an extraterrestrial civilization utilizing the total energy of the star; and Type-III is an advanced alien society, that consumes the total galactic energy. Many aspects of the mentioned societies have been considered in a series of papers \citep{DSBH,DSWD,type1,paper1,paper2,dyson}, but in this paper, we consider only galactic civilizations, thus, Type-I and Type-II societies and study the possibility of how visible we are to them. In particular, the question is: can the artifacts of our technological society be visible and potentially detectable by the telescopes of ETs. Since the question is to identify our society with civilization, the major focus should be on the search for large ships, buildings and space satellites etc. Such artifacts might easily be identified as artificial constructions.
For this purpose, it is natural to focus on the visible light reflected from the corresponding objects. To identify an observed object with an artificial one, the best way is to spatially resolve it. Therefore, optical telescopes will be used. But, on the other hand, the angular resolution depends on the diameter of a telescope, $\theta\simeq 1.22\times \lambda/D$ \citep{carroll}, which for very small angles requires extremely large telescopes (here, $\lambda$ is the wavelength). Instead of using large telescopes of astronomical sizes (although, such a possibility cannot be excluded from consideration), one can apply long baseline optical interferometry \citep{interfer}, by using at least two telescopes separated by a huge distance - baseline, $B$, when the resolving angle might be very small, $\theta\simeq\lambda/B$. The processing of data might be significantly simplified by the possibility of recording the optical signal,  which is now  impossible because no device can record a signal with a time-scale of the order of $10^{-15}-10^{-14}$ sec, a typical scale of the corresponding period. In \citep{dvali} we have shown that Type-I,II,III advanced alien societies might use quantum computers based on artificial black holes, which are able to record the mentioned signals, since the minimum time-scales might be much less than $10^{-15}-10^{-14}$ sec.

In this paper, we analyze how visible we are to advanced ETs, depending on their technological level.

The article is organized as follows: In Sec. 2, we
consider the basic ideas and obtain the results, and in Sec. 3, we summarize them.

\section{Discussion and results}

In order to understand the capabilities of advanced civilizations, it is better to first estimate the timescale for reaching the corresponding technological levels. As we have already mentioned, the Type-I alien society is the one that utilizes the total power incident from the solar-type host star on an Earth-type planet, $P_{_{I}}\simeq \frac{L_{\odot}}{4}\times\left(\frac{R_E}{R_{AU}}\right)^2\simeq 1.7\times 10^{24}$ erg s$^{-1}$, where $L_{\odot}\simeq 3.8\times 10^{33}$ erg s$^{-1}$ denotes the Solar luminosity, $R_{AU}\simeq 1.5\times 10^{13}$ cm is the distance from the Sun to the Earth (one astronomical unit - AU) and $R_E\simeq 6.4\times 10^{8}$ cm is the Earth's radius. Unlike the mentioned power, our current power consumption, $1.5\times 10^{20}$ erg s$^{-1}$, is by several orders of magnitude less. By following \cite{dyson} and assuming that $1\%$ of annual growth rate of industry is maintained, one can straightforwardly show that our civilization, which is of type $0.75$, will reach Type-I in approximately $t_I = 1000$ yrs. For the Type-II the corresponding time-scale equals $t_{II} = 3000$ yrs (see also \cite{dyson}). Therefore, it is natural to assume that advanced alien societies might be able to launch telescopes on circular trajectories separated by the distance of the order of $\sim 10 AU$. \footnote{Even our civilization was able to launch Voyager-I,II, which have already moved away from the sun at distances respectively $160$ AU and $130$ AU: https://voyager.jpl.nasa.gov/.}

Then, by considering an object of a length-scale, $l$, which should be spatially resolved at the maximum distance $r_{max}$ from it, one can obtain an expression of $r_{max}$ by combining two expressions of the angular resolution, $\theta\simeq\lambda/B$ and $\theta\simeq l/r_{max}$
\begin{equation}
\label{r} 
r_{max}\simeq B\times \frac{l}{\lambda}\simeq 3000\times\frac{B}{10\;AU}\times\frac{l}{10\;m}\times\frac{550nm}{\lambda}\; ly.
\end{equation}
Here we have taken into account that the wavelength for visible light lies in the interval $(400-700)$ nm and we used its average value, $550$ nm. The estimate shows that ETs will be able to identify artificial constructions from distances of $\sim 3$ kly. For example, the pyramids of Egypt might have been detected. On the other hand, the artificial satellites can be visible, but if the observing ETs are more than $60$ ly away, detection will be impossible because the optical signal from the first artificial satellites has traveled only $\sim 60$ ly.

In order to be sure, that the advanced societies can detect our constructions one should study the flux sensitivity of the observed objects. For this purpose, it is necessary is to estimate the spectral luminosity of the solar-type stars in the visible area \citep{carroll}
\begin{equation}
\label{Llam} 
L_{vis}\simeq 4\pi R_{\odot}^2\times\frac{2\left(kT\right)^4}{h^3c^2}\times\int_{x_{min}}^{x_{max}}\frac{x^3}{e^x-1}dx\simeq 4.5\times 10^{32} \; erg/s,
\end{equation}
where $T\simeq 5777$ K is the temperature of the star's surface, $R_{\odot}\simeq 7\times 10^{10}$ cm is its radius, $k$ is the Boltzmann's constant, $c$ is the speed of light, $h$ denotes the Planck's constant, $x_{min}=\epsilon_{min}/(kT)$, $x_{max}=\epsilon_{max}/(kT)$ and $\epsilon_{min} = hc/\lambda_{max}$ and $\epsilon_{max}=hc/\lambda_{min}$ are respectively the minimum and maximum photon energies corresponding to the visible spectrum ($\lambda_{min} = 400$ nm, $\lambda_{max} = 700$ nm). Then, if one intends to observe an object of the length-scale, $l$, reflecting a fraction $\xi$ of the incident light of a star from a distance, $R$, visible with the angle, $\theta_0$, for the power incident on the telescope one obtains
\begin{equation}
\label{pow} 
P\simeq\xi\times l^2\times\frac{L_{vis}}{4\pi R^2}\times\frac{\theta_0^2}{2\pi}.
\end{equation}
For the observation time-scale, $\tau$, the total energy $P\times\tau$ should be guaranteed by the incident $N$ photons with the average energy, $\epsilon=hc/\lambda$. Therefore, the energy balance condition, $P\times\tau\simeq\epsilon N$ combined with Eq. (\ref{r}) and with a natural condition, $\theta_0\simeq D/r_{max}$ leads to an estimate of the telescope diameter
$$D\simeq\frac{2RB}{\lambda}\times\left(\frac{\pi hcN}{\xi\lambda\tau L_{vis}}\right)^{1/2}\simeq$$
\begin{equation}
\label{D} 
\simeq 3\times 10^6\times\frac{R}{1\;AU}\times\frac{B}{10\; AU}\times\left(\frac{550}{\lambda}\right)^{3/2}\times\left(\frac{N}{10^6}\times\frac{0.5}{\xi}\times\frac{1\; hour}{\tau}\right)^{1/2}\; km,
\end{equation}
where we have assumed that the reflection coefficient equals $0.5$, the typical distance between a star and a planet and the baseline are of the order of $1$ AU and $10$ AU respectively, the time-scale of observation is of the order of $1$ hour, and the number of photons required to resolve a building should be at least $10^6$.

From Eq. (\ref{D}) it is clear that the diameter of the telescope should be on the order of several million kilometers. Such huge megastructures might be built only by Type-II civilizations, but not by Type-I alien societies. In particular, in \citep{type1} it has been shown that during a typical time-scale, $\tau_I=1000$ yrs, one might be able to construct only earth-sized megastructures. Therefore, henceforth, the focus will be on Type-II advanced technologies.

Analyzing the maximum distances, it is significant to estimate the possible distribution of advanced ETs in the Milky Way galaxy.

In his seminal work \cite{drake} has derived an expression to estimate the number of communicating civilizations
\begin{equation}
\label{drake} 
N = R_{\star}\times f_p\times n_e\times f_l\times f_i\times f_t\times \mathcal{L},
\end{equation}
where $R_{\star}$ is the rate of star formation in the galaxy, $f_p$ is the fraction of stars with planetary systems, $n_e$ denotes the average number of habitable planets per star, $f_l$ is the fraction of planets potentially supporting life which actually developed it, $f_i$ is the fraction of planets with intelligent life, $f_t$ denotes the fraction of technological civilizations that can communicate, and $\mathcal{L}$ denotes the average length of a time-scale for which ET civilizations release signals into space. 

From modern observations, the value of the star formation rate is relatively well defined $R_{\star} = (1.5-3)$ stars per year \citep{rate}. The Kepler mission has enabled us to estimate that the average number of planets in the habitable zones is on the order of $40$ billion. According to this study $f_p\times n_e\simeq 0.4$ \citep{fpne}. Since life emerged on Earth soon after the conditions became favorable for life, it is widely accepted that $f\sim 1$. According to the statistical approach developed by \cite{maccone} its value is close to $1/3$ and $f_i\times f_t$ is of the order of $0.01$. In general, it is believed that $f_l\times f_i\times f_t$ should vary in the interval $(10^{-3}-1)$. Following the discussion presented in \citep{dvali} and assuming $\mathcal{L}_{II}\gtrsim t_{II}$, one can  estimate an interval of the number of Type-II civilizations: $N_{II} = (6-3.6\times 10^3)$. It is worth noting that the maximum values are derived for $\mathcal{L}_{II}\simeq t_{II}$ and, consequently, it is clear that if the civilizations exist for longer time-scales, the values of $N_{II}$ might be (significantly) increased.

As an order of magnitude, we assume that the civilizations are uniformly distributed over the galactic plane. Then, considering the upper limit of $N_{II}$ and taking an average diameter of MW into account $D_{MW}\simeq 26,8$ kpc \citep{carroll}, for an average distance among civilizations one obtains
\begin{equation}
\label{dist2} 
R_{II}\simeq\left(\frac{\pi D_{MW}^2}{4N_{II}}\right)^{1/2}\simeq 1300 \; ly.
\end{equation}
As one can see from the obtained value, $r_{max}>R_{II}$, indicating that technologically advanced ETs might detect our large constructions, belonging to the period from ancient up to medieval times. One can straightforwardly check that if the number of civilizations is not less than $650$ our civilization will be visible to them.

They can detect our modern constructions only if their total number in the MW is of the order of $10^6$, which has been hypothesized by \cite{sagan}. This is possible if the civilization time-scale is of the order of $10^6$ yrs. In this context it is worth noting that star ages in the MW might differ from the Solar age by starting from hundreds of millions of years up to several billions of years \citep{carroll}, therefore, if evolution has started in an older planetary system, the time-scale of the technological civilization might be quite large.

\section{Summary}

In the paper we have considered the possibility of detection of our technological artifacts by advanced civilizations by using optical interferometry and mega-telescopes.

We have found that the maximum distance, where a $10$ meter length-scale construction might be spatially resolved is of the order of $3000$ ly.

By analysing the spectral sensitivity, it has been shown that the telescope diameters might be extremely large for Type-I civilizations to be constructed and only Type-II advanced societies might be able to spatially resolve the artificial constructions.

By analysing the Drake equation, it has been found that if the number of civilizations is of the order of $650$ they will be able to detect our artificial constructions.

\section{Acknowledgments}
The research was partially supported by the EU fellowships for Georgian researchers, 2023 (57655523). Z.O. also would like to thank Torino Astrophysical Observatory and Universit\'a degli Studi di Torino for hospitality during working on this project.

\end{document}